\documentclass[aps,showpacs,amsmath,amssymbd,dvips,showkeys,preprint,preprintnumbers]{revtex4}

\usepackage{graphicx}
\usepackage{color}

\def \be  {\begin{equation}}
\def \ee  {\end{equation}}
\def \ee  {\end{equation}}
\def \bea {\begin{eqnarray}}
\def \eea {\end{eqnarray}}

\begin{document}

\preprint{ECTP-2017-04}
\preprint{WLCAPP-2017-04}
\hspace{0.05cm}
\title{Phenomenology of light- and strange-quark simultaneous production at high energies}

\author{Abdel Nasser  Tawfik}
\email{a.tawfik@eng.mti.edu.eg}
\affiliation{Egyptian Center for Theoretical Physics (ECTP), Modern University for Technology and Information (MTI), 11571 Cairo, Egypt}
\affiliation{World Laboratory for Cosmology And Particle Physics (WLCAPP), 11571 Cairo, Egypt}

\author{Hayam Yassin}
\author{Eman R. Abo Elyazeed}
\affiliation{Physics Department, Faculty of Women for Arts, Science and Education, Ain Shams University, 11577 Cairo, Egypt}

\date{\today}

\begin{abstract}
This letter presents an extension of EPL116(2017)62001 to light- and strange-quark nonequilibrium chemical phase-space occupancy factors ($\gamma_{q,s}$). The resulting damped trigonometric functionalities relating $\gamma_{q,s}$ to the nucleon-nucleon center-of-mass energies $(\sqrt{s_{NN}})$ looks very similar except different coefficients. The phenomenology of the resulting $\gamma_{q,s}(\sqrt{s_{NN}})$ describes a rapid decrease at $\sqrt{s_{NN}}\lesssim7~$GeV followed by a faster increase up to $\sim20~$GeV. Then, both $\gamma_{q,s}$ become nonsensitive to $\sqrt{s_{NN}}$. Although these differ from $\gamma_{s}(\sqrt{s_{NN}})$ obtained at $\gamma_q(\sqrt{s_{NN}})=1$, various particle ratios including $\mathrm{K}^+/\pi^+$, $\mathrm{K}^-/\pi^-$, $\mathrm{\Lambda}/\pi^-$, $\bar{\mathrm{\Lambda}}/\pi^-$, $\mathrm{\Xi}^+/\pi^+$, and $\mathrm{\Omega}/\pi^-$, can well be reproduced, as well. We conclude that $\gamma_{q,s}(\sqrt{s_{NN}})$ should be instead determined from fits of various particle yields and ratios but not merely from fits to the particle ratio $\mathrm{K}^+/\pi^+$. 
\end{abstract}


\maketitle


\section{Introduction} 
\label{intro}

The present letter is devoted to the phenomenology of both light- and strange-quark chemical phase-space occupancy factors as observed in a wide range of the nucleon-nucleon center-of-mass energies ($\sqrt{s_{NN}}$) \cite{Tawfik:2017ehz}.  It is assumed that the final state production of the hadron yields is conjectured to be sensitive to the chemical phase-space available to the constituent light- and strange-quarks, which can be an equilibrium ($\gamma_q=\gamma_s=1$) or a full nonequilibrium ($\gamma_q\neq1$ and $\gamma_s\neq1$) of even a partly nonequilibrium ($\gamma_q\neq1$ or $\gamma_s\neq1$) process. The latter was analyzed in Ref.  \cite{Tawfik:2017ehz}. The chemical phase-space occupancy factors ($\gamma_{q,s}$) determine the number of produced pairs of a certain quark flavor. Accordingly, the final state particle-antiparticle pairs can be regulated. Furthermore, $\gamma$'s are conjectured to manifest whether the process of the light- and strange-quark production is proceeded in or out of equilibrium. 

The number of the final state hadron yields is dictated by the quarks fugacities [$\lambda_{q,s}^{(i)}=\prod_{k\in i}\exp(\mu_k/T)$] and their chemical phase-space occupancy factors ($\gamma_{q,s}$), i.e. $N_i\sim \exp(\varepsilon_i/T)\prod_{k\in i}\lambda_k \gamma_k$, where the $i$-th particle dispersion relation reads $\varepsilon_i=(m_i^2+p_i^2)^{1/2}$ \cite{RL2001}. We assume that both up- and down-quarks are degenerate, i.e. $\gamma_q\equiv\gamma_u=\gamma_d$ and $\mu_q\equiv\mu_u=\mu_d$. Accordingly, there is light-quark isospin symmetry, i.e. vanishing isospin potential ($\mu _{I_3}$). 

Nonequilibrium chemical phase-space occupancy for the strange quark $\gamma_{s}\neq1$ was introduced in order to characterize the possible strangeness enhancement \cite{Rgmm1}, which in turn was assumed to be resulted in from the formation of the quark-gluon plasma (QGP), the new state-of-matter discovered at RHIC energies \cite{Gyulassy:2004zy} and later on confirmed though systematic jet-quenching analysis at LHC energies \cite{Betz:2012hv,Angerami:2012tr}. Here, we assume that both chemical phase-space occupancy factors $\gamma_{q,s}\neq1$ \cite{Tawfik:2017ehz,Tawfik:2014eba,Tawfik:2005gk}. 

In additional to $T$, $\mu_b$, and $\gamma_{q,s}$, another flow parameter would play an important role, especially when certain particle species are taken into consideration.  The flow parameter can be determined through integrating the entire chemical phase-space. Alternatively, the longitudinal flow would be enough, when the yield multiplicities are  detected per unit of rapidity. The impact of this parameter shall be analyzed in a future work. 

The characteristic Marek's {\it  horn-like} structure \cite{Marek1999} was confirmed in various high-energy experiments \cite{Expp1,Expp2,Expp3}. There is an increase in $\mathrm{K}^+/\pi^+$ with increasing $\sqrt{s_{NN}}$ up to top SPS energies \cite{Expp4}). A further increase in $\sqrt{s_{NN}}$ is accompanied by a rapid decrease in $\mathrm{K}^+/\pi^+$. Then, $\mathrm{K}^+/\pi^+$ becomes nonsensitive to a further increase in the energies. At LHC energies, another puzzle was observed that the proton-to-pion similar to the kaon-to-pion ratios are overestimated by the thermal models. This observation contradicts the well-established success of the statistical thermal models in describing various aspects of the particle production, especially the particle yields and ratios, at a wide range of  $\sqrt{s_{NN}}$ \cite{Tawfik:2014eba}. 

The success reported in Ref. \cite{Tawfik:2017ehz}, where $\gamma_q$ was {\it ad hoc} assumed to remain in equilibrium while merely $\gamma_q\neq1$, in reproducing various particle ratios encourages us to assume full chemical nonequilibrium, i.e. $\gamma_s\neq1$ and $\gamma_q\neq1$. The procedure we follow is a statistical fit of the horn-like structure of $\mathrm{K}^+/\pi^+$, where $\gamma_q$ and $\gamma_s$ are added to the temperature ($T$) and the baryon chemical potential ($\mu_b$) as free parameters, Fig. \ref{fig:1}.  The resulting $\gamma_q$ and $\gamma_s$ shall be utilized in the partition function in order to calculate further particle ratios.

\section{Approach}
\label{sec:apprch}

The Laplace transform of the accessible phase space characterized by $\gamma_q$ and $\gamma_s$ results in an expression looks mathematically similar to the partition function based on the Hagedorn approach, e.g. the hadron resonance gas (HRG) model, \cite{RL2001}. 
\bea
{\ln Z}\left(T,\mu\right) &=&\pm \frac{V}{2{\pi }^2} \sum_i g_i  \int^{\infty }_0
k^2\, dk  \ln  \left[1 \pm \left(\gamma_q^{n_q}\right)_i\, \left(\gamma_s^{n_s}\right)_i \, 
  e^{\left(\frac{\mu_i-\varepsilon_{i}(k)}{T}\right)}\right], \label{eq:lnZ} 
\eea
where $\pm$ stands for fermions and bosons, respectively, and $g_i$ is the degeneracy factor of $i$-the hadron resonance. The chemical potential ($\mu_i$) is assumed to count for various contributions, for instance, 
\begin{equation} 
\mu_i=B_{i} \mu _{B} + S_{i} \mu _{S} +  I_{3_i} \mu _{I_3} + \cdots,
\end{equation}
where $V$ is the fireball volume and $B_i$, $S_i$, and  $I_{3_i}$ are baryon, strangeness, and isospin quantum number of $i$-th hadron resonance and  $\mu_{B}$, $\mu_{S}$, and $\mu _{I_3}$ are the baryon, strangeness, and isospin chemical potential, respectively.  $n_q$ and $n_s$ are the number of light- and strange-quarks, respectively, of which such fermion or boson is composed. 

\begin{figure}[htb]
\centering{
\includegraphics[width=5.75cm,angle=-90]{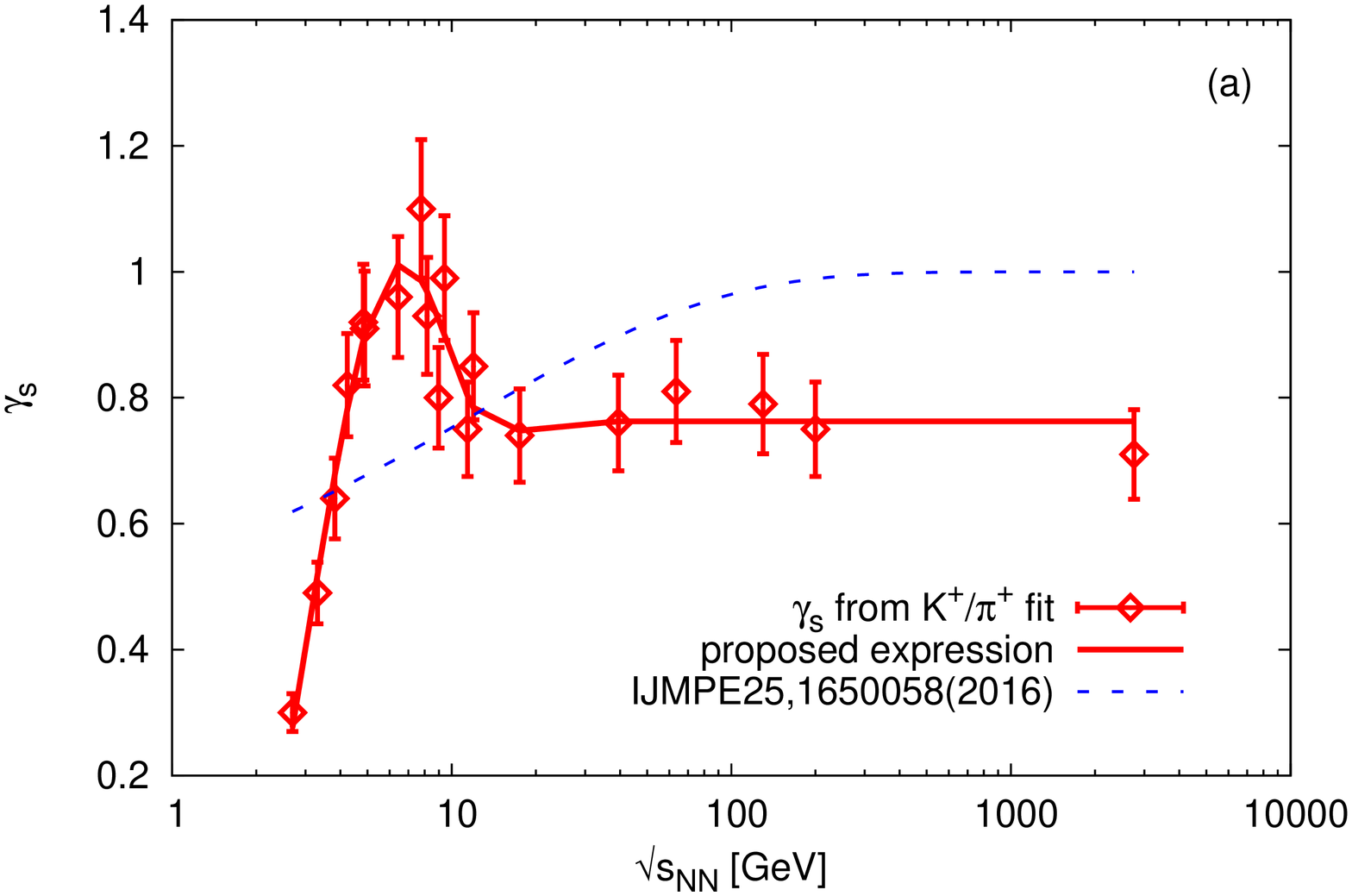}
\includegraphics[width=5.75cm,angle=-90]{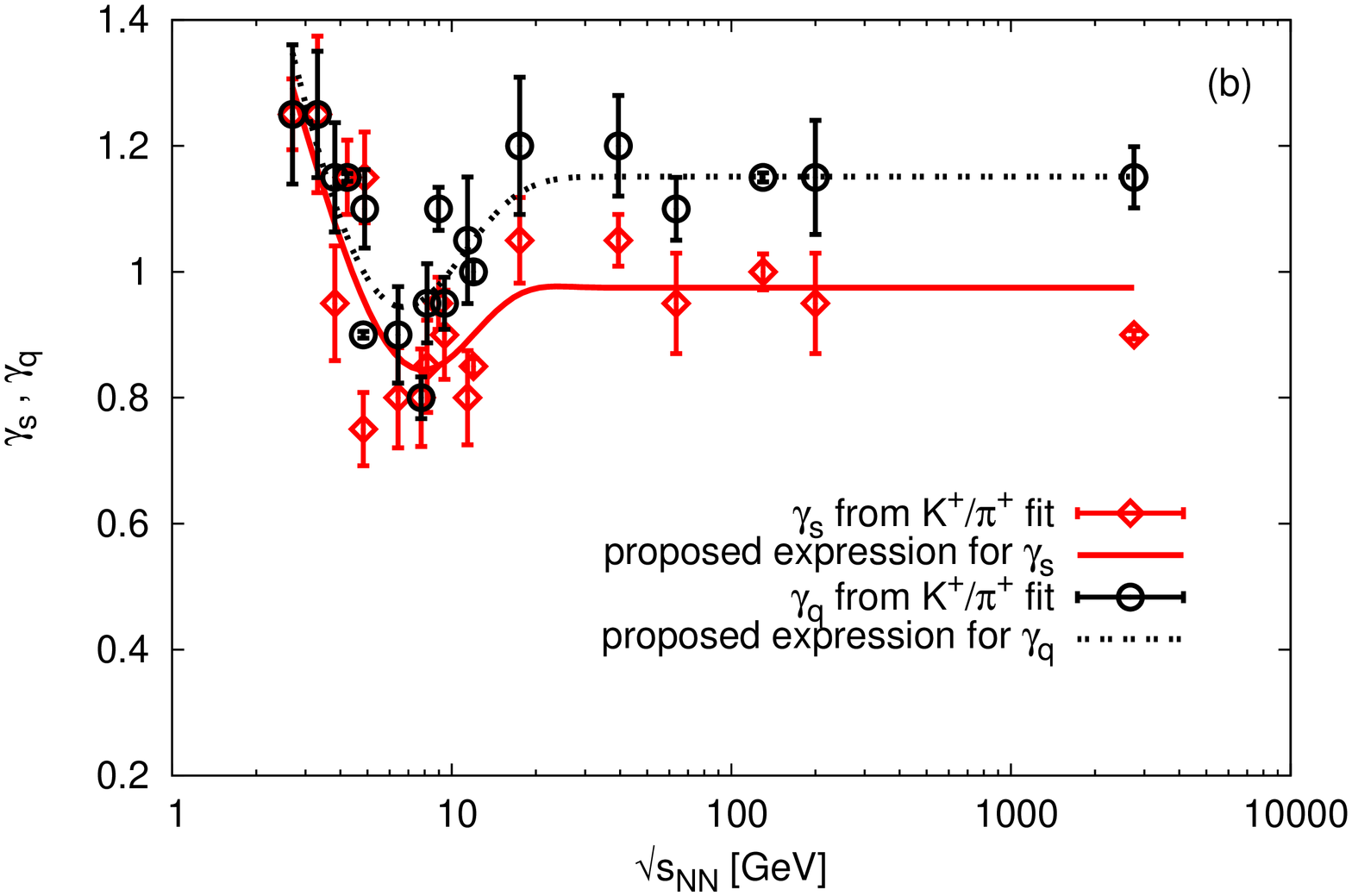}
\caption{(Color online)  Resulting $\gamma_q$ and $\gamma_s$ (symbols) are given in dependence on $\sqrt{s_{NN}}$. The curves represent the proposed damped  trigonometric functionalities.
\label{fig:1}
}}
\end{figure}

The energy dependence of the $\mathrm{K}^+/\pi^+$ ratios calculated from the HRG model poorly reproduce the horn-like structure measured, especially at top SPS energies. This overestimates the measurements at higher $\sqrt{s_{NN}}$, as well, especially at $\gamma_q=\gamma_s=1$ \cite{Tawfik:2014eba,Tawfik:2005gk}. In the present letter, we assume that both chemical phase-space occupancy factors are free parameters enabling the HRG calculations to fit well the measured $\mathrm{K}^+/\pi^+$ \cite{Tawfik:2017ehz,Tawfik:2005gk}. The resulting $\gamma_q$ and $\gamma_s$ are depicted in left-hand panel of Fig. \ref{fig:1}. 

At nonequilibrium $\gamma_q$ and $\gamma_s$ \cite{Rgmm1}, the statistical nature of light- and strange-quark production is assumed to be maintained. In the present work, we present results based on the same procedure. The measured $\mathrm{K}^+/\pi^+$ ratios are fitted to the HRG calculations, where the freezeout parameters (the temperature and the baryon chemical potential and accordingly other types of potentials such as strangeness and isospin) are determined (fixed) at $s/T^3=7$ \cite{Tawfik:2004ss,Tawfik:2005qn}, with $s$ is the entropy density. Both sets of $\mathrm{K}^+/\pi^+$ ratios (measured and calculated) almost excellently approach each other, at varying $\gamma_q$ and $\gamma_s$.

The resulting $\gamma_q$ and $\gamma_s$ have an opposite qualitative nonmonotonic dependence on $\sqrt{s_{NN}}$ as that of the $\mathrm{K}^+/\pi^+$ ratios. The energy dependence of both $\gamma_q$ and $\gamma_s$ is formulated as damped  trigonometric functionalities, 
\bea
\gamma_{q,s}(\sqrt{s_{NN}}) &=& a_{q,s}\, \exp\left(-b_{q,s}\, \sqrt{s_{NN}}\right)\, \sin\left(c_{q,s}\, \sqrt{s_{NN}} + d_{q,s}\right) + f_{q,s},\label{eq:Fitgmqs} 
\eea
where $a_q=-6.138\pm0.837$, $b_q=0.308\pm 0.062$, $c_q=0.085\pm0.004$, $d_q=5.978\pm1.041$, and $f_q=1.151\pm0.036$ and $a_s=-1.733\pm1.768$, $b_s=0.262\pm 0.062$, $c_s=0.196\pm0.047$, $d_s=5.373\pm1.044$, and $f_s=0.975\pm0.047$. Both expressions fits well with the resulting $\gamma_q$ and $\gamma_s$, Fig. \ref{fig:1} (b). 


We first recall the $\gamma_{s}(\sqrt{s_{NN}})$ functionality at $\gamma_{q}(\sqrt{s_{NN}})=1$ \cite{Tawfik:2017ehz}, left-hand panel of Fig. \ref{fig:1}. At AGS energies, HRG calculations, for instance, at {\it equilibrium} chemical phase-space occupancy factors reproduce well the measured $\mathrm{K}^+/\pi^+$ ratios. So do the {\it partly} nonequilibrium calculations \cite{Tawfik:2017ehz}. Furthermore, we highlight that the insensitivity of $\gamma_s(\sqrt{s_{NN}})$ to the hadron resonance masses and to the excluded volume corrections \cite{Tawfik:2005gk,Satz2016}. Different from Ref. \cite{Tawfik:2017ehz}, the present analysis shows a rapid suppression in the light- and strange-quark simultaneous production, at AGS energies, Fig. \ref{fig:1}. With increasing $\sqrt{s_{NN}}$, $\gamma_s$ exponentially increases. This is reversed at $7 \lesssim \sqrt{s_{NN}}\lesssim 20~$GeV. At higher energies, the energy dependence vanishes.  

The present parameterizations, Eqs. (\ref{eq:Fitgmqs}), keep only the asymptotic energy-independent region, at high energies. At energies lower than to SPS, an opposite $\sqrt{s_{NN}}$-dependence is obtained. Both $\gamma_q$ and $\gamma_s$ decrease with increasing $\sqrt{s_{NN}}$. At $\sqrt{s_{NN}}\simeq 7~$GeV, both reach minimum values. This is exactly in opposition to what was reported on in Ref. \cite{Tawfik:2017ehz}, which is characterized by an {\it ad hoc} $\gamma_q=1$. A further increase in $\sqrt{s_{NN}}$ is then accompanied with increasing $\gamma_q$ and $\gamma_s$. At $\sqrt{s_{NN}}\gtrsim20~$GeV, both asymptotic energy-independent values set on, i.e. $\gamma_{q,s}=f_{q,s}$, Fig. \ref{fig:1} (b). 

The crucial question to be answered now is why both parameterizations, Eqs. (\ref{eq:Fitgmqs}), and the previous one \cite{Tawfik:2017ehz}, substantially differ at low energies, although they are mathematically identical? In the section that follows, we shall discuss on the differences between both types of parameterizations and their physical consequences. We shall analyze how the varying coefficients/parameters contribute to the various energy-dependencies? While the implications of the previous parameterization were limited to reproduction of various particle ratios \cite{Tawfik:2017ehz}, the present ones shall be also applied in reproducing various particle ratios including $\mathrm{K}^+/\pi^+$, $\mathrm{K}^-/\pi^-$, $\mathrm{\Lambda}/\pi^-$, $\bar{\mathrm{\Lambda}}/\pi^-$, $\mathrm{\Xi}^+/\pi^+$, and $\mathrm{\Omega}/\pi^-$. Other implications shall be presented, as well.
 
\section{Particle Yields and Ratios }

\begin{figure}[htb]
\centering{
\includegraphics[width=3.75cm,angle=-90]{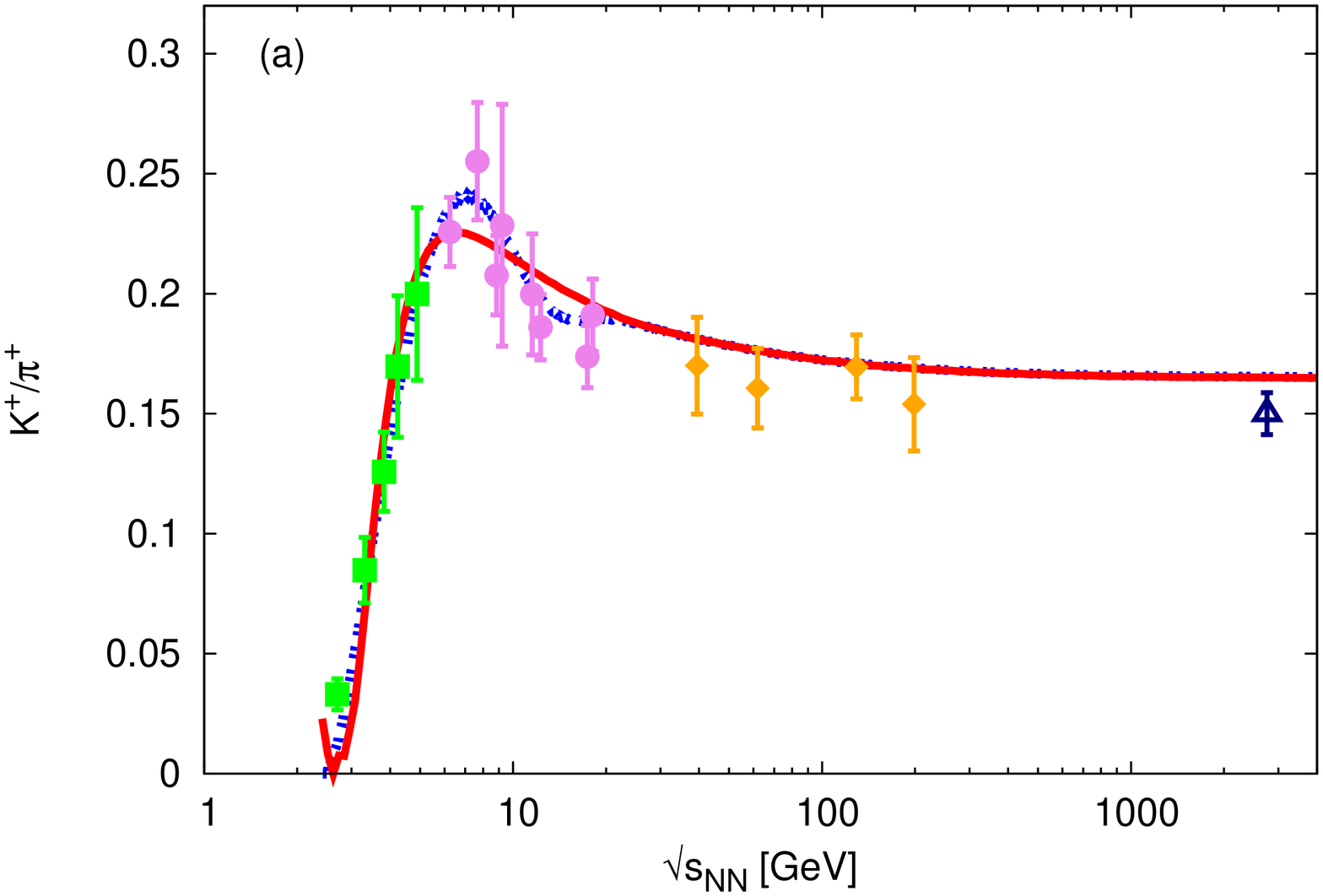}
\includegraphics[width=3.75cm,angle=-90]{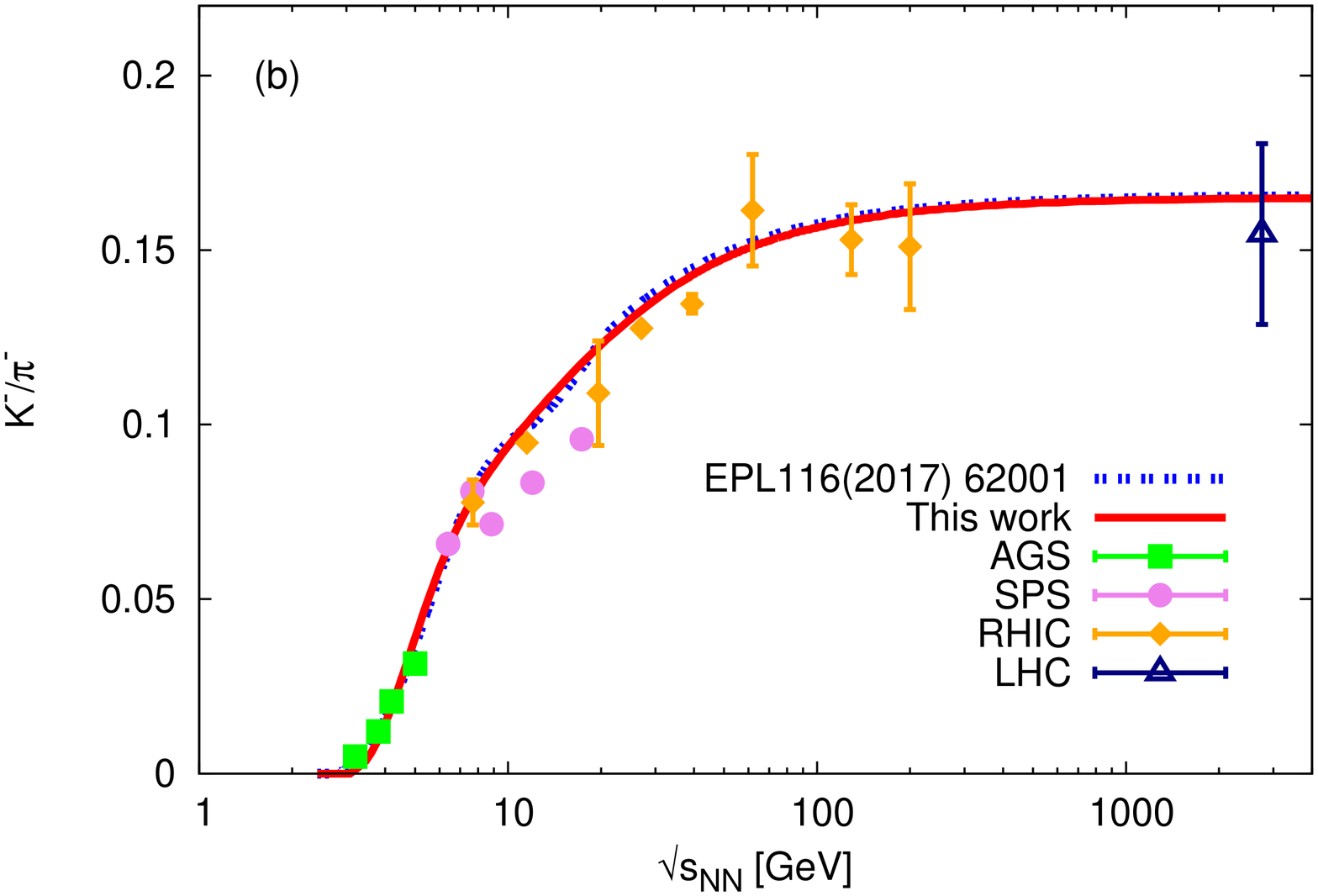} 
\includegraphics[width=3.75cm,angle=-90]{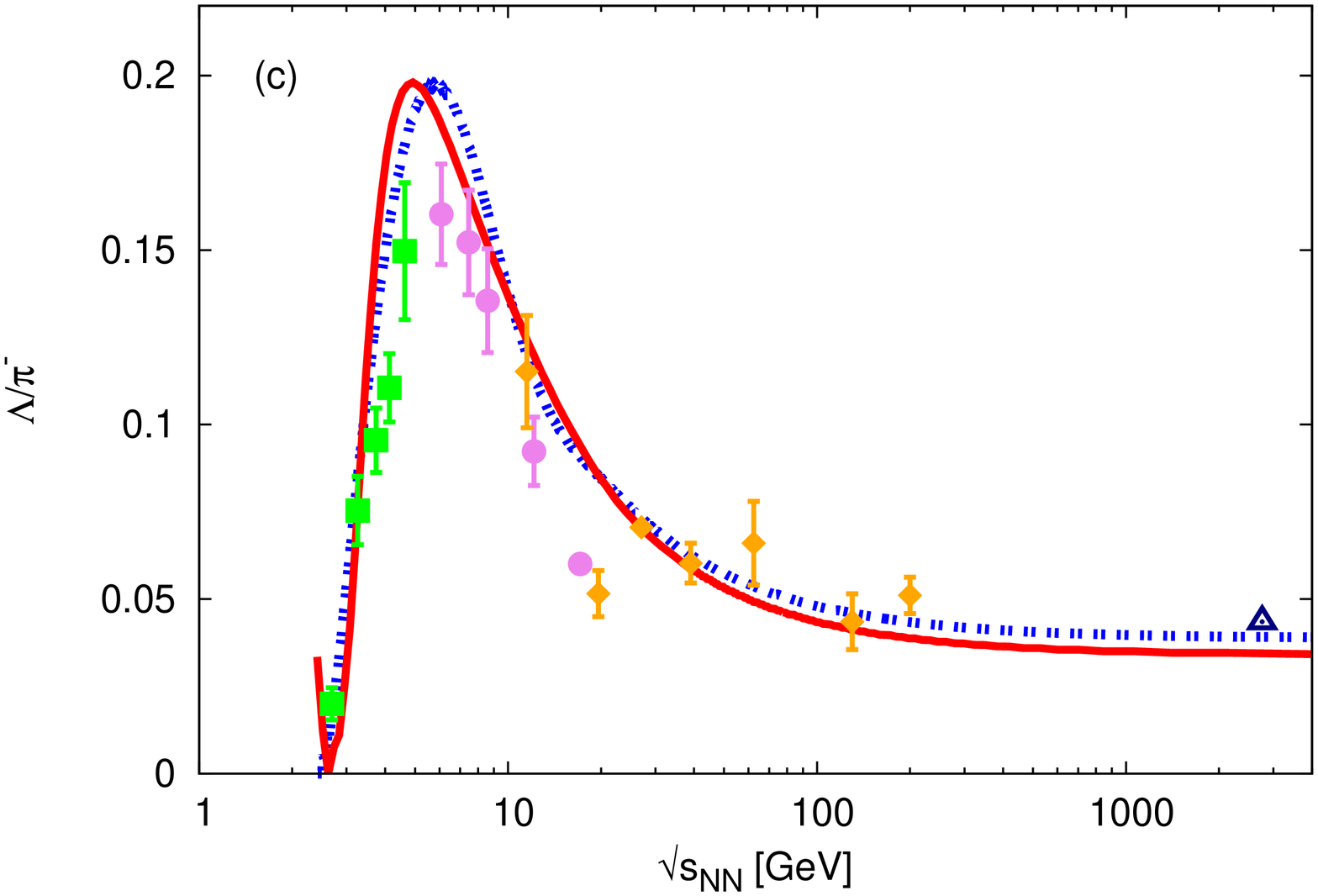}\\
\includegraphics[width=3.75cm,angle=-90]{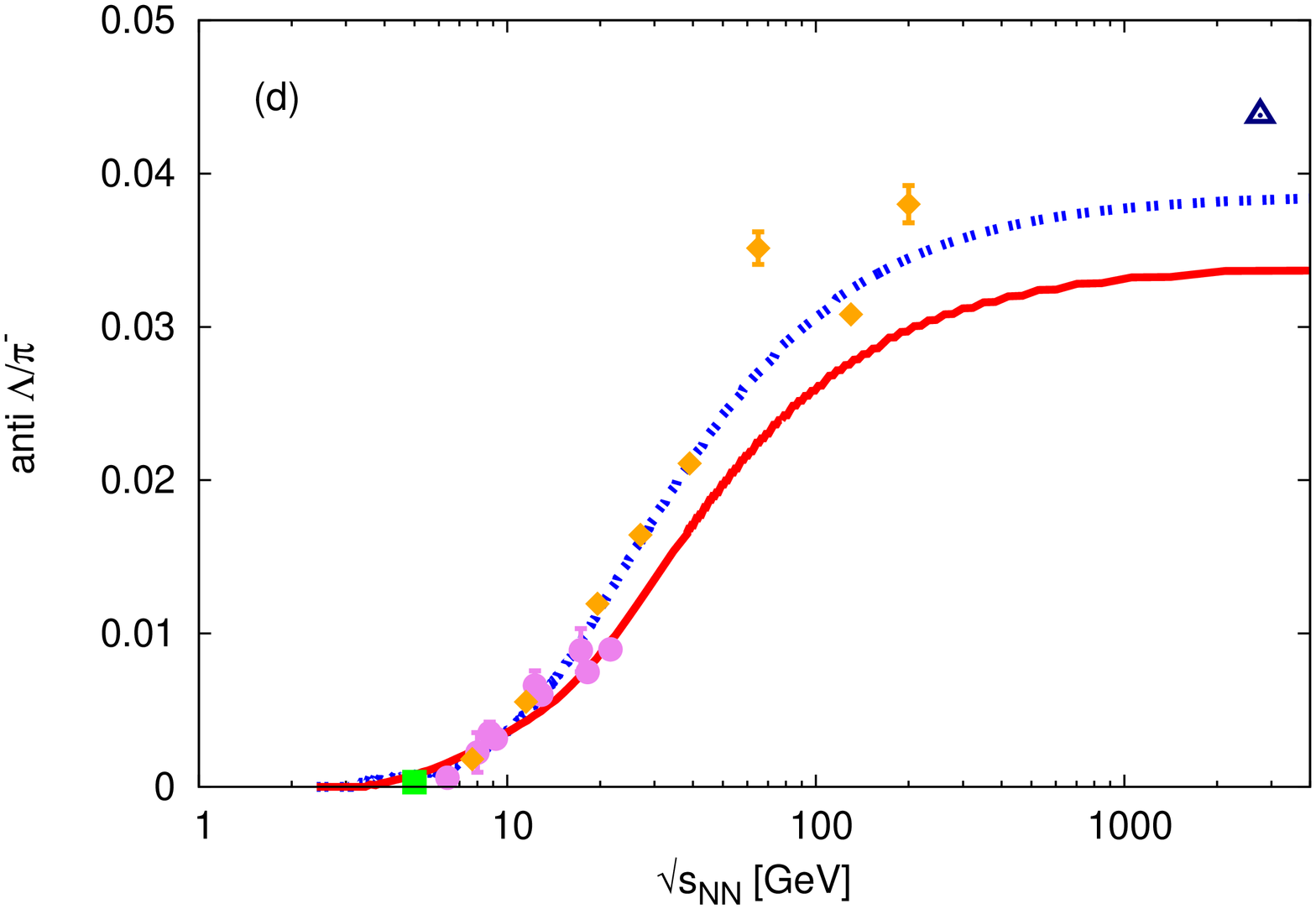} 
\includegraphics[width=3.75cm,angle=-90]{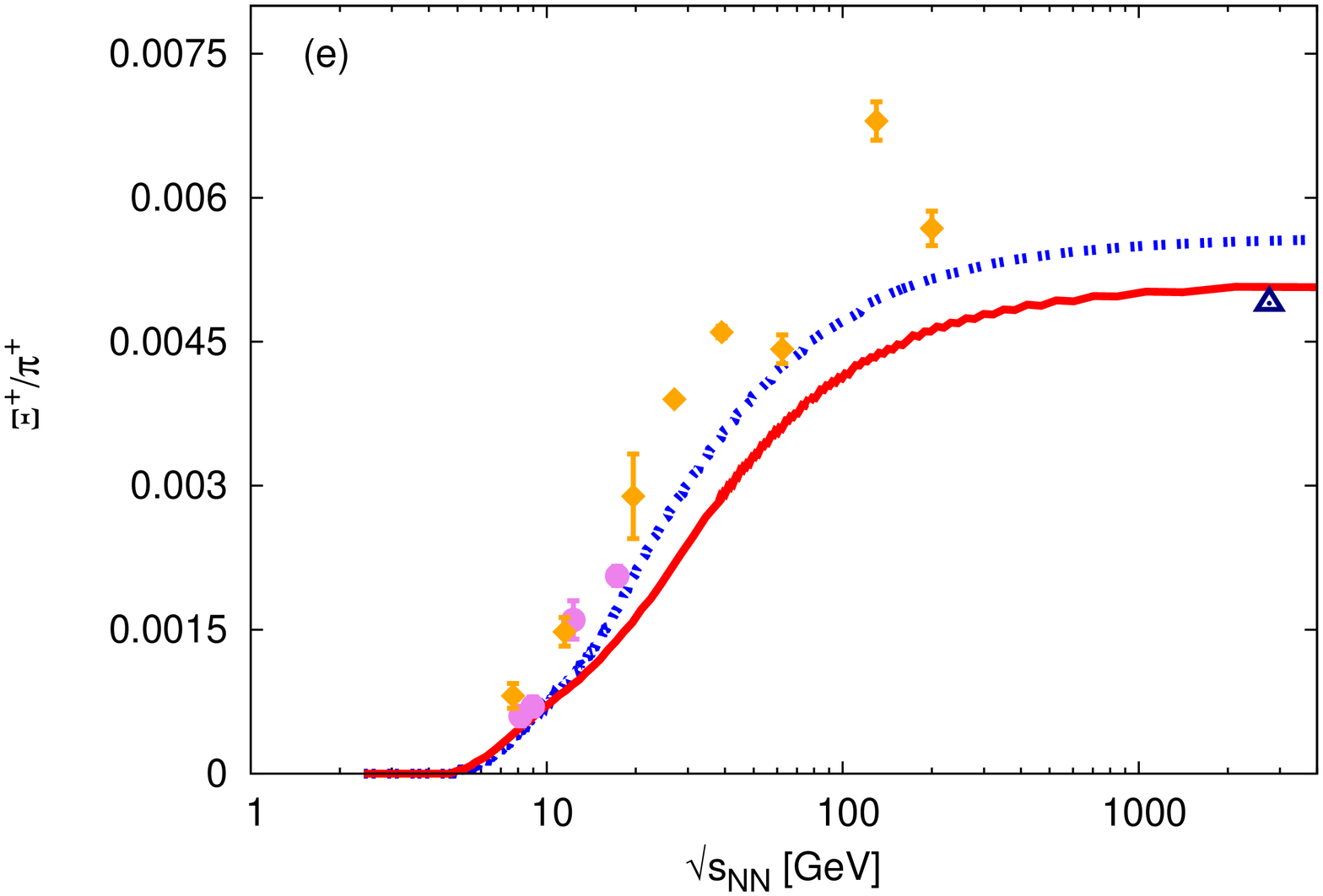}
\includegraphics[width=3.75cm,angle=-90]{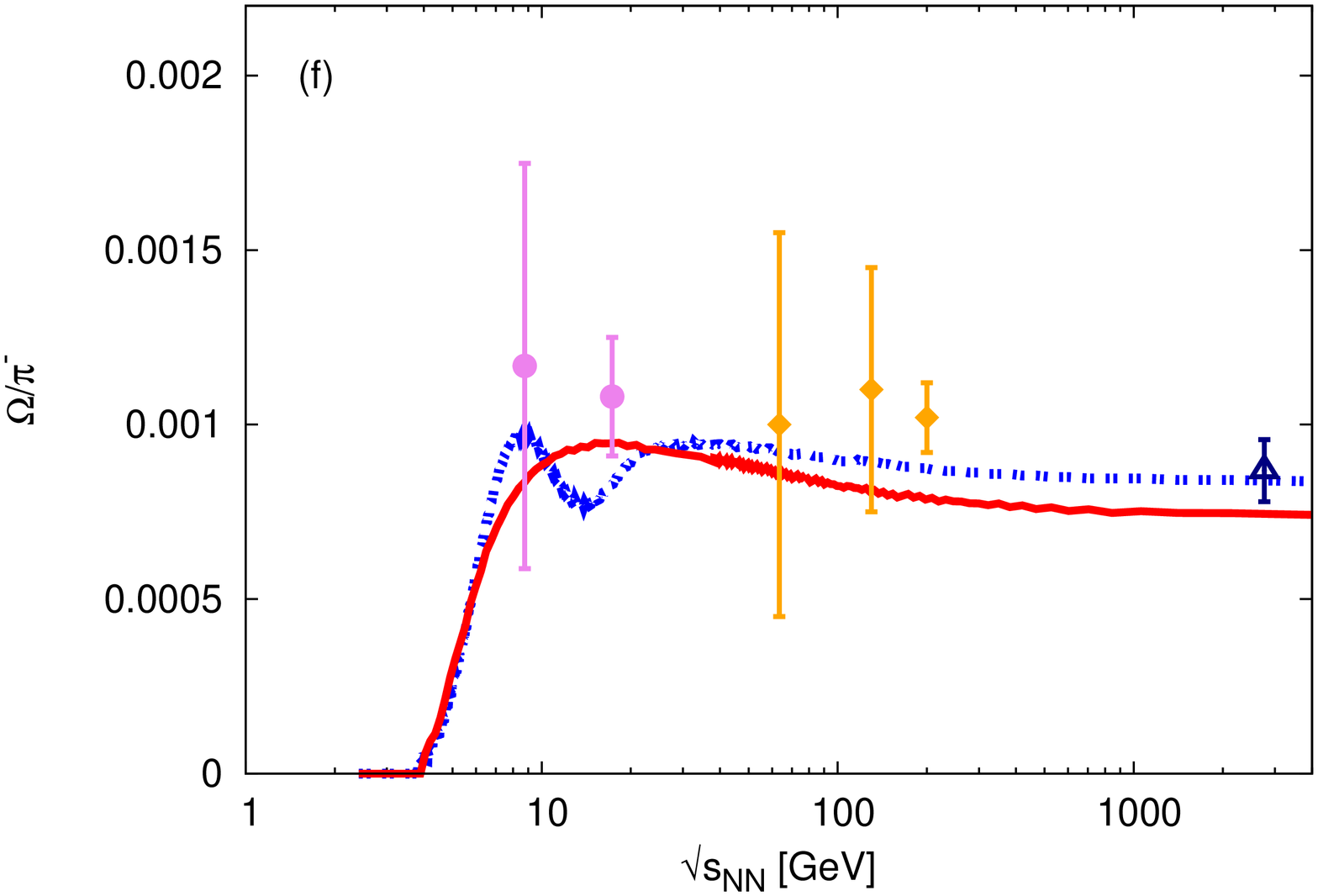}
\caption{(Color online)  The energy dependence of measured $\mathrm{K}^+/\pi^+$ (a), $\mathrm{K}^-/\pi^-$ (b), $\mathrm{\Lambda}/\pi^-$ (c), $\bar{\mathrm{\Lambda}}/\pi^-$ (d), $\mathrm{\Xi}^+/\pi^+$ (e), and $\mathrm{\Omega}/\pi^-$ (f) is compared with the HRG calculations at partly (double-dotted curves) \cite{Tawfik:2017ehz} and full nonequilibrium chemical phase-space occupancy factors (solid curves).
\label{fig:2}
}}
\end{figure} 

Figure \ref{fig:2} shows a systematic comparison between various particle ratios including  $\mathrm{K}^+/\pi^+$ (a), $\mathrm{K}^-/\pi^-$ (b), $\mathrm{\Lambda}/\pi^-$ (c), $\bar{\mathrm{\Lambda}}/\pi^-$ (d), $\mathrm{\Xi}^+/\pi^+$ (e), and $\mathrm{\Omega}/\pi^-$ (f) measured in a wide range of energies and the HRG calculations (symbols with errorbars) at partly (double-dotted curves) and full nonequilibrium \cite{Tawfik:2017ehz} chemical phase-space occupancy factors $\gamma_{q,s}$ (solid curves). We notice that the HRG calculations for $\mathrm{K}^+/\pi^+$ (a), $\mathrm{K}^-/\pi^-$ (b), $\mathrm{\Lambda}/\pi^-$ (c), and $\mathrm{\Omega}/\pi^-$ (f) are improved when assuming full nonequilibrium chemical phase-space occupancy, Eq. (\ref{eq:Fitgmqs}), while $\bar{\mathrm{\Lambda}}/\pi^-$ (d) and $\mathrm{\Xi}^+/\pi^+$ (e) are slightly made worse. This is also valid when comparing partly with full nonequilibrium $\gamma_{q,s}$. 

The $\gamma_{q,s}(\sqrt{s_{NN}})$ deduced from the statistical fits of multiple-strange-quarks systems are also parameterized. We found that their implications (not shown here) does not differ from the one deduced from the single-strange-quark system $\mathrm{K}^+/\pi^+$.

We conclude that the parameterizations $\gamma_{q,s}(\sqrt{s_{NN}})$ bring remarkable improvements to the ability of the thermal models to reproduce various particle ratios. We also conclude that these parameterizations works very well in some particle ratios but unfortunately not in others. 

\begin{figure}[htb]
\centering{
\includegraphics[width=5.5cm,angle=-0]{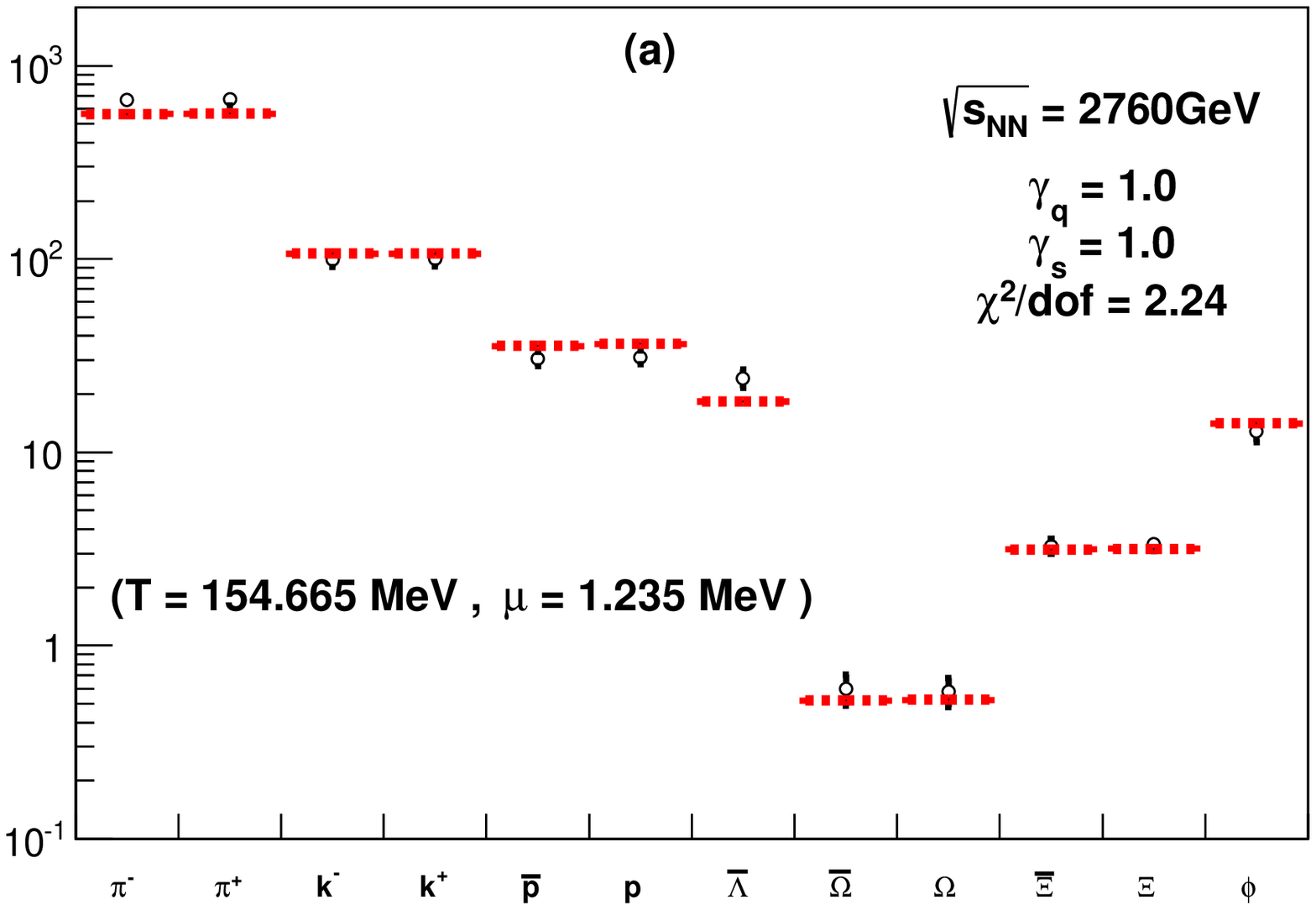}
\includegraphics[width=5.5cm,angle=-0]{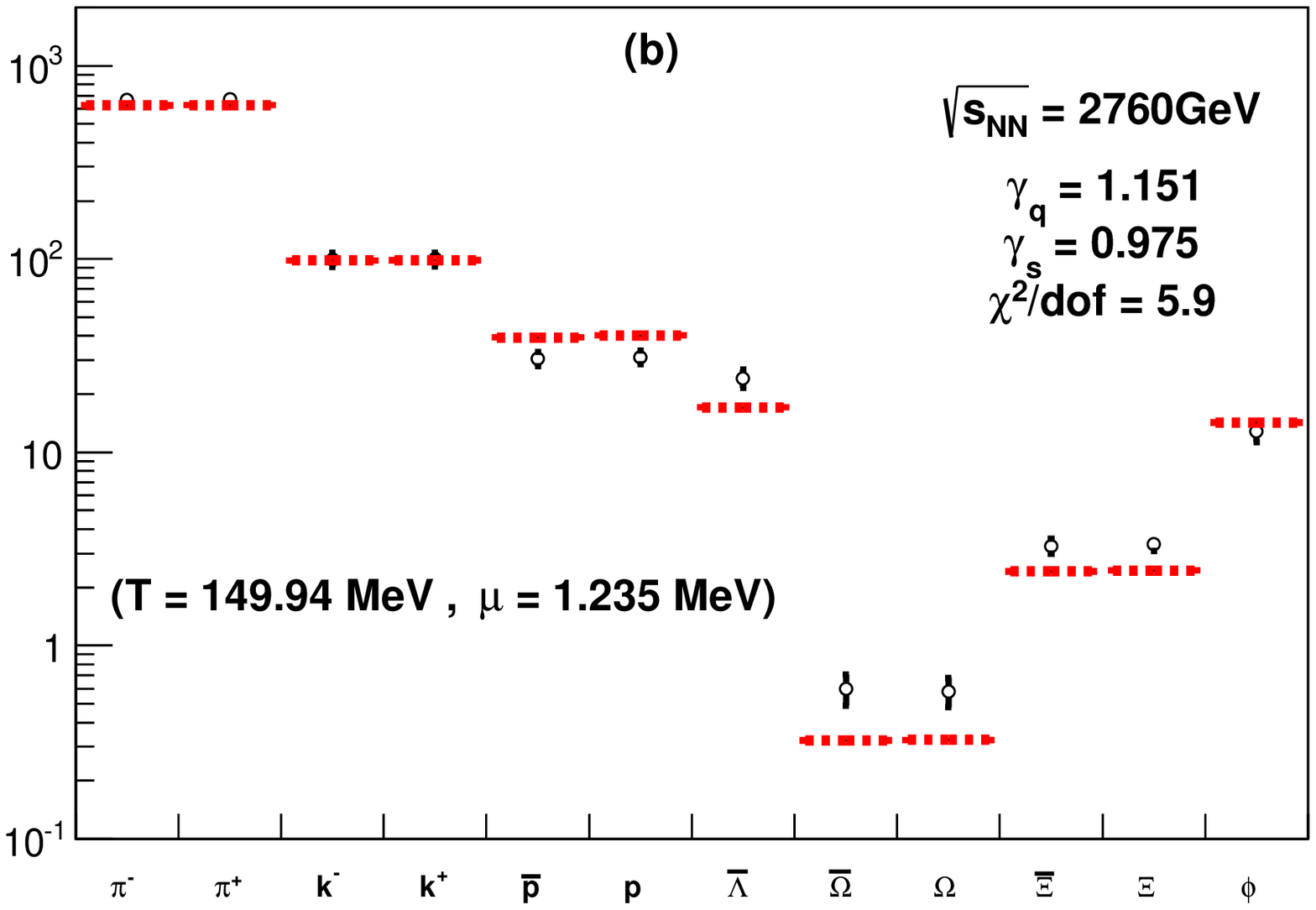}
\includegraphics[width=5.5cm,angle=-0]{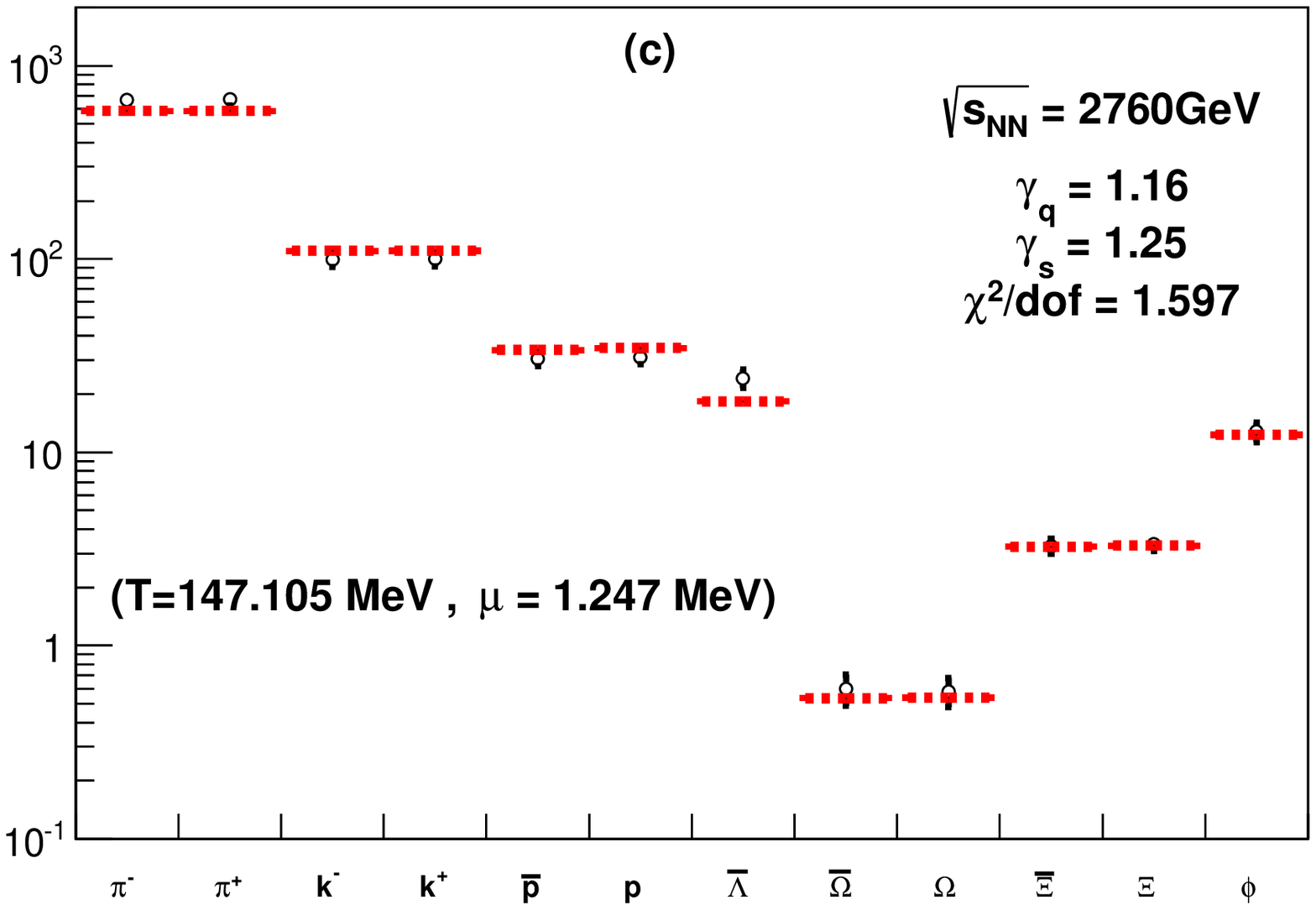}\\
\includegraphics[width=5.5cm,angle=-0]{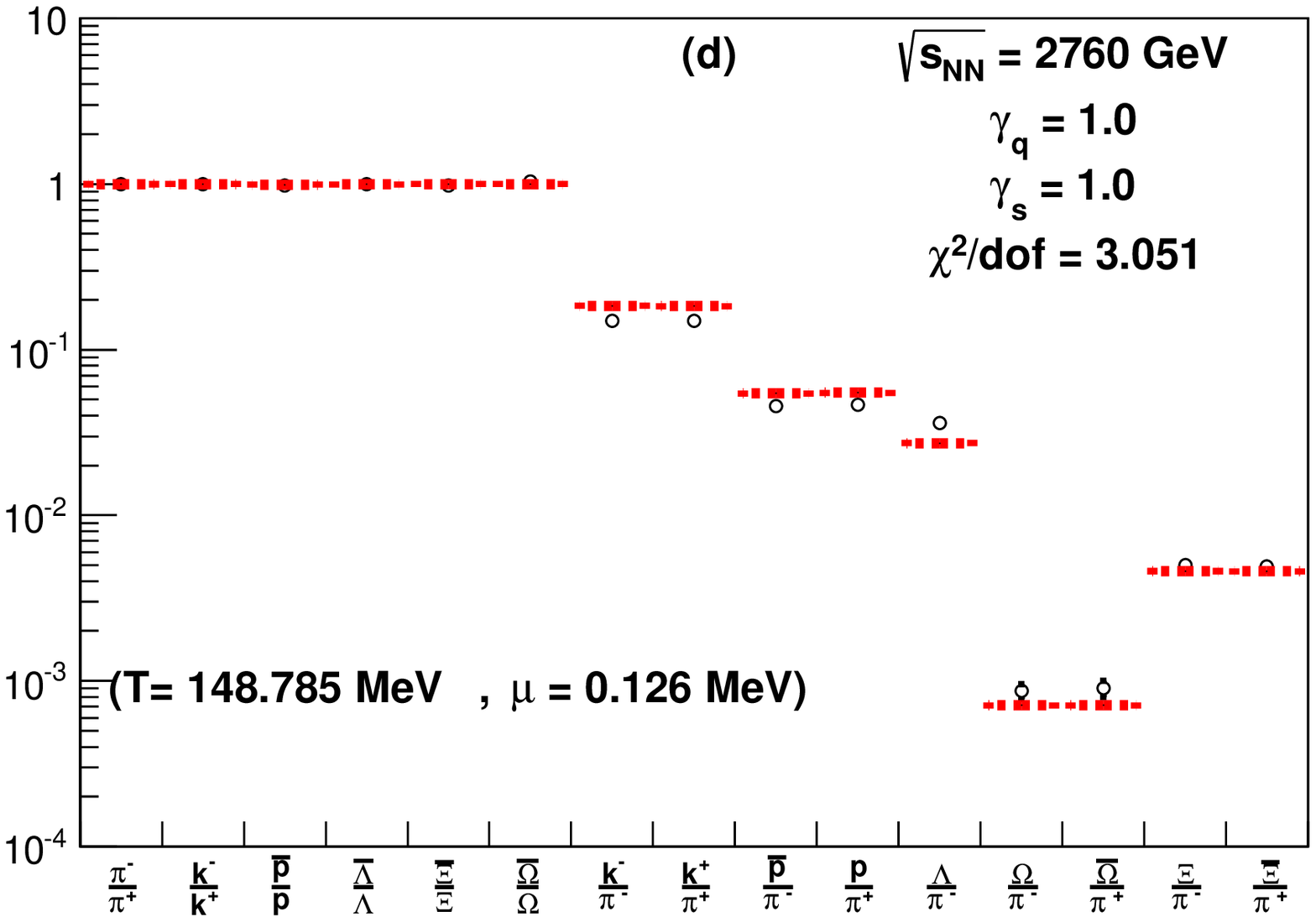}
\includegraphics[width=5.5cm,angle=-0]{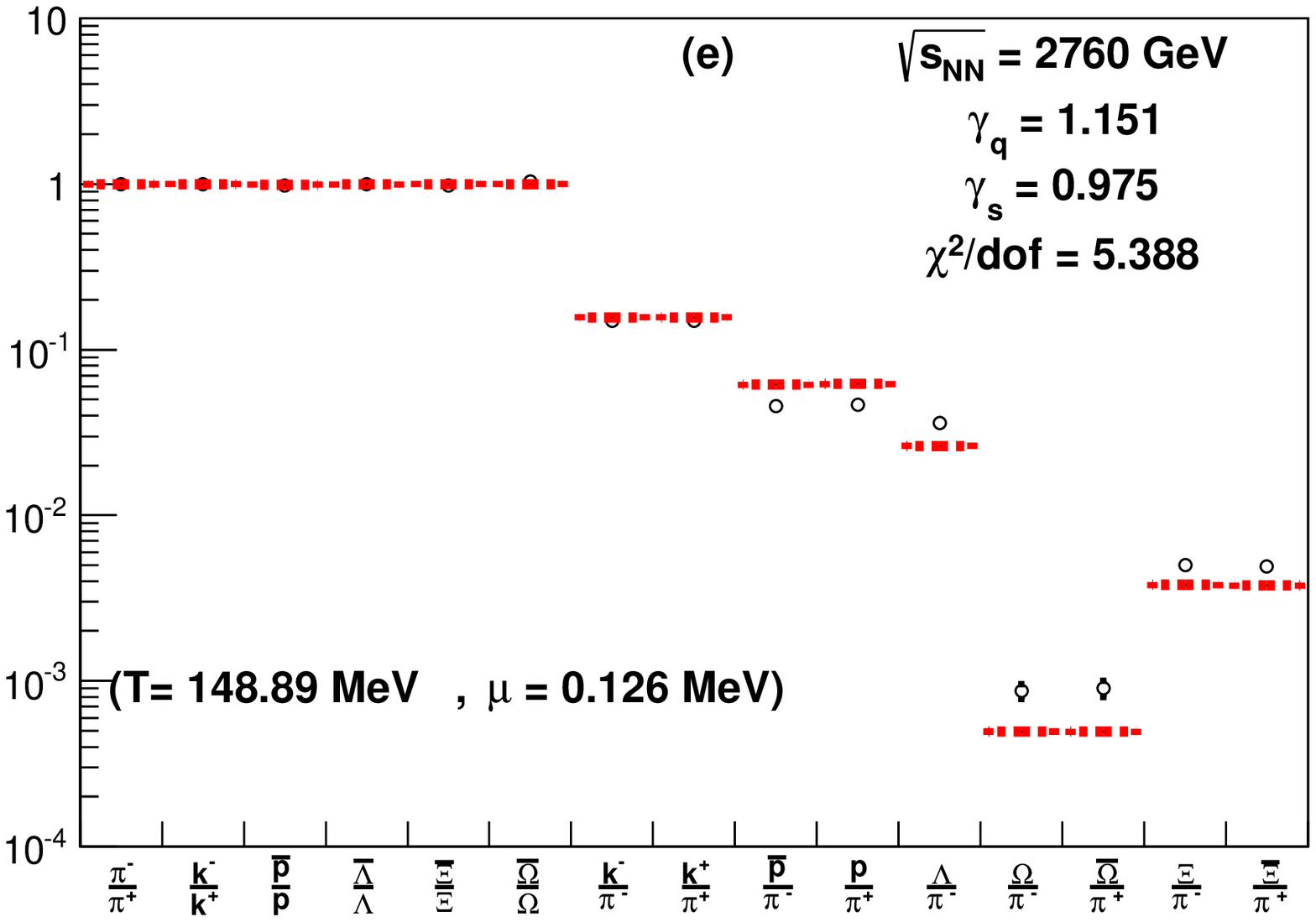}
\includegraphics[width=5.5cm,angle=-0]{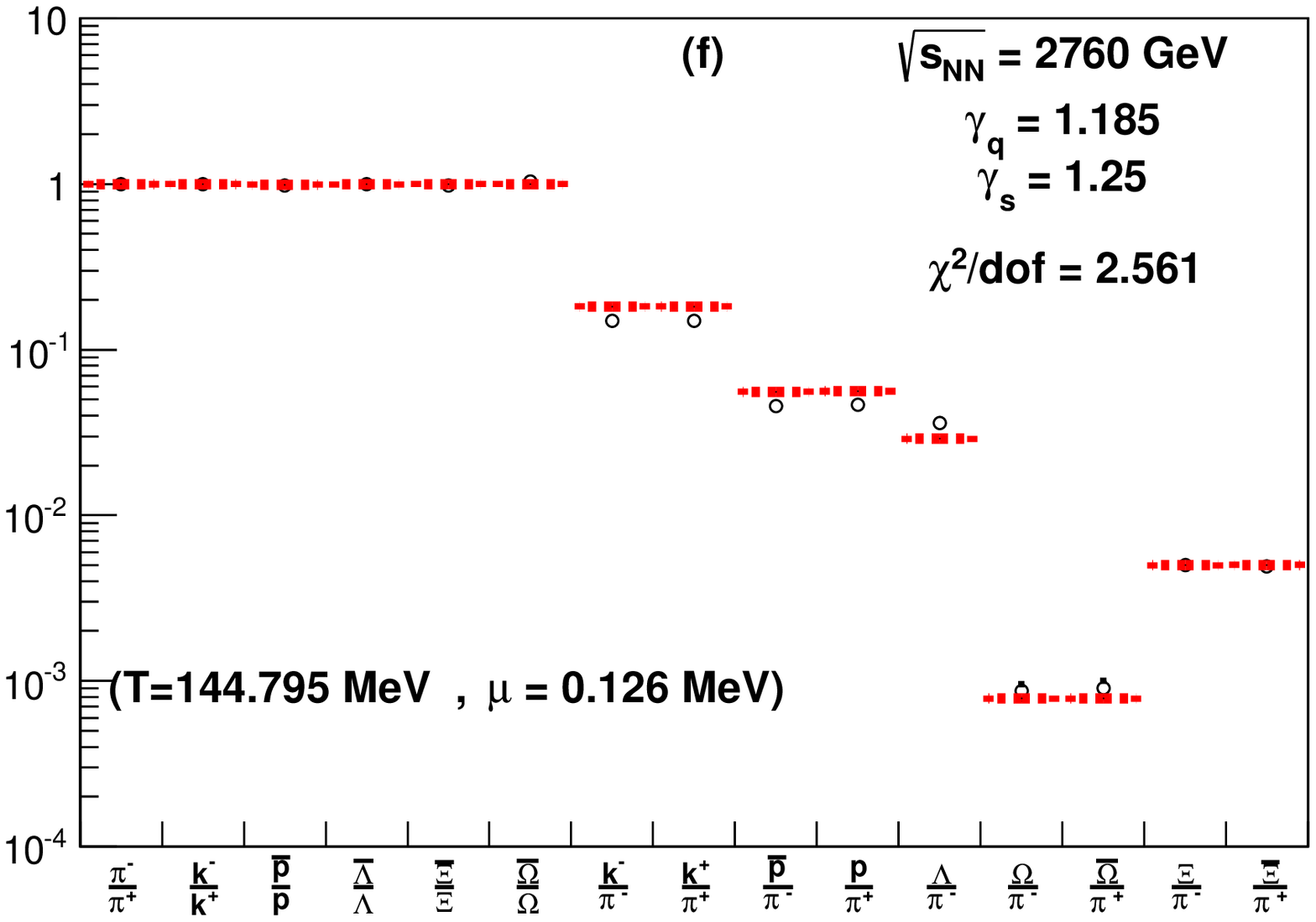}
\caption{(Color online) Various particle yields (top panel) and ratios (bottom panel) are fitted through HRG with equilibrium ($\gamma_{q,s}=1$) and full nonequilibrium ($\gamma_{q,s}\neq1$) chemical phase-space occupancy factors. 
\label{fig:3}
}}
\end{figure}

In order to analyze how good $\gamma_{q,s}(\sqrt{s_{NN}})$ reproduce various particle yields and ratios, we focus on  results at $\sqrt{s_{NN}}=2760~$GeV, for instance.  Fig. \ref{fig:3} depicts different particle yields (top panel) and ratios (bottom panel) measured in central $Pb+Pb$ collisions at $2760~$GeV (symbols) and compares these to HRG calculations at equilibrium $\gamma_{q,s}=1$ (left-hand panel), full nonequilibrium $\gamma_{q,s}=f_{q,s}$ (middle and right-hand panels) chemical phase-space occupancy factors. The results at full nonequilibrium distinguish between calculations at $\gamma_{q,s}=f_{q,s}$ (middle panel), where $\gamma_{q,s}$ are fitted to $\mathrm{K}^+/\pi^+$, the present work and the ones, where $\gamma_{q,s}$ are adjusted from the best reproduction of various particle yields and ratios, simultaneously, i.e. additional statistical fits. The goodness of the latter is better than the earlier (compare corresponding $\chi^2/$dof). As observed in Fig. \ref{fig:2}, $\gamma_{q,s}=f_{q,s}$ based on fit of the $\mathrm{K}^+/\pi^+$ ratio reproduce well most of particle ratios. Some of them are over- or underestimated. This can be interpreted due to possible limitation of the single-strange-quark system $\mathrm{K}^+/\pi^+$. 

The right-hand panel presents an excellent reproduction of various particle yields and ratios, where the freezeout parameters are well determined; $T=147.105~$MeV, $\mu_b=1.247~$MeV (particle yields) and $T=144.795~$MeV, $\mu_b=0.126~$MeV (particle ratios). When comparing the results at $\gamma_{q,s}=f_{q,s}$ (middle panel) with the ones at varying $\gamma_{q,s}$, we find that  $\gamma_{q,s}=f_{q,s}$ are modified to $\gamma_q=1.16$ and $\gamma_s=1.25$ (particle yields) and to $\gamma_q=1.185$ and $\gamma_s=1.25$ (particle ratios). The reproduction of various particle yields and ratios is remarkable improved (compare corresponding $\chi^1/$dof). The statistical fits for the particle yields are accompanied by an estimation for the fireball volume or radius; \\
at $\gamma_{q,s}=1$, $V=20442.24~$GeV$^{-3}$ or $R=3.392~$fm, \\
at $\gamma_{q,s}=f_{q,s}$,  $V=20367.9~$GeV$^{-3}$ or $R=3.388~$fm, and \\
at $\gamma_{q}=1.185$ and $\gamma_{s}=1.25$, $V=20301.12~$GeV$^{-3}$ or $R=3.384~$fm.

%
\section{Conclusions} 
\label{conclusion}

We have presented a systematic analysis for the light- and strange-quark nonequilibrium chemical phase-space occupancy factors ($\gamma_{q,s}$) based on statistical fit of the $\mathrm{K}^+/\pi^+$ ratios measured in a wide range of energies. The freezeout condition $s/T^3=7$ \cite{Tawfik:2004ss,Tawfik:2005qn}, where $s$ is the entropy density, is utilized in order to determine the freezeout parameters, the temperature and the baryon chemical potential, while $\gamma_{q,s}$ are taken as free parameters. Resulting $\gamma_{q,s}$ are parameterized as damped trigonometric functionalities in $\sqrt{s_{NN}}$. They are mathematically identical to $\gamma_{s}(\sqrt{s_{NN}})$ obtained at $\gamma_q(\sqrt{s_{NN}})=1$ \cite{Tawfik:2017ehz}. We notice that varying the various coefficients/parameters substantially change their nonmonotonic behaviors, especially at low energies. While $\gamma_{s}(\sqrt{s_{NN}})$ at $\gamma_q(\sqrt{s_{NN}})=1$ increases at low energies, $\gamma_{q,s}(\sqrt{s_{NN}})$ decrease. Also, when $\gamma_{s}(\sqrt{s_{NN}})$ at $\gamma_q(\sqrt{s_{NN}})=1$ reaches maximum at low energies, $\gamma_{q,s}(\sqrt{s_{NN}})$ reach minimum. But both sets of  parameterizations are qualitatively identical, at high energies. Before we discuss on the possible explanation of the resulting $\gamma_{q,s}(\sqrt{s_{NN}})$ relative to $\gamma_{s}(\sqrt{s_{NN}})$ at $\gamma_q(\sqrt{s_{NN}})=1$, we first highlight that various particle ratios including $\mathrm{K}^+/\pi^+$, $\mathrm{K}^-/\pi^-$, $\mathrm{\Lambda}/\pi^-$, $\bar{\mathrm{\Lambda}}/\pi^-$, $\mathrm{\Xi}^+/\pi^+$, and $\mathrm{\Omega}/\pi^-$ are well  reproduced, as well, while others not.

The observation that both parameterizations enable the hadron resonance gas model to reproduce well various particle ratios, although their different energy-dependencies, at low energies, suggests that both degrees of freedom (light and strange quarks) are strongly correlated with each others. The suppression in one of them seems to be accompanied by a suppression in the other one and vice versa. Such a serious conclusion needs an unambiguous confirmation preferably from the first-principle lattice QCD simulations and/or alternatively from the perturbation theory. The scope of the present work is a phenomenological study for the characteristic $\mathrm{K}^+/\pi^+$ ratios. We first complete (remain within) this given frame. We recall that the implementation of the resulting $\gamma_{q,s}(\sqrt{s_{NN}})$ in reproducing various particle ratios shows that $\mathrm{K}^+/\pi^+$, $\mathrm{K}^-/\pi^-$, $\mathrm{\Lambda}/\pi^-$, and $\mathrm{\Omega}/\pi^-$ are well, while $\bar{\mathrm{\Lambda}}/\pi^-$ and $\mathrm{\Xi}^+/\pi^+$ are slightly worse reproduced. 

For a fair assessment of the resulting $\gamma_{q,s}(\sqrt{s_{NN}})$, we focus the discussion on  results at a given energy, $\sqrt{s_{NN}}=2760~$GeV. At equilibrium ($\gamma_{q,s}=1$), full nonequilibrium ($\gamma_{q,s}=f_{q,s}$) chemical phase-space occupancy factors, different particle yields and ratios are compared with the HRG calculations. The full nonequilibrium distinguishes between calculations at $\gamma_{q,s}=f_{q,s}$, where $\gamma_{q,s}$ are merely fitted to $\mathrm{K}^+/\pi^+$, and free $\gamma_{q,s}$. The latter assures a best reproduction of the various particle yields and ratios, simultaneously. In other words, $\gamma_{q,s}$ are fitted at this given energy.  

Differ from $f_{q,s}$, $\gamma_q=1.16$ and $\gamma_s=1.25$ for particle yields and $\gamma_q=1.185$ and $\gamma_s=1.25$ for particle ratios, enable the HRG model to excellently reproduce the available experimental results. Accordingly, the resulting freezeout parameters are as follows. 
$T=147.105~$MeV and $\mu_b=1.247~$MeV from fits of particle yields and $T=144.795~$MeV and $\mu_b=0.126~$MeV from fits of particle ratios. An estimation for the fireball volume or radius is also obtained $V=20301.12~$GeV$^{-3}$ or $R=3.384~$fm.
 
In light of this, we conclude that the reliable estimation of $\gamma_{q,s}$ shouldn't be limited to a statistical fit of single-strange-quark system, such as $\mathrm{K}^+/\pi^+$ or even just one particle ratio. We have to extend this to various particle yields and ratios. Other precise measurements such as flow, transverse momentum distributions and rapidity multiplicities would also be taken into account. The guess that the production (suppression or enhancement) of the light quarks should be correlated with the production of strange quarks can first be confirmed if this shall remain valid when extending the statistical fits. To such a systematic analysis, we shall devote a future work.


\end{document}